\documentclass[10pt]{amsart}

\usepackage{amsmath}
\usepackage{braket}
\usepackage{graphicx}
\usepackage{subfigure}
\usepackage{ulem}
\usepackage{cite}

\title{On the role of interactions in trans-sonically flowing atomic condensates}
\author{Manuele Tettamanti$^{1,2}$ \and Iacopo Carusotto$^{3}$ \and Alberto Parola$^{4}$}

\address{\noindent $^1$ipartimento di Fisica ``Giuseppe Occhialini'', Universit\`{a} di Milano-Bicocca - Piazza della Scienza 3, 20126 Milano, Italy \endgraf
$^2$INFN -- Sezione di Milano-Bicocca - Piazza della Scienza 3, 20126 Milano, Italy \endgraf
$^3$INO-CNR BEC Center and Dipartimento di Fisica, Universit\`{a} di Trento - I-38123 Povo, Italy \endgraf
$^4$Dipartimento di Scienza e Alta Tecnologia, Universit\`{a} degli Studi dell'Insubria - Via Valleggio 11, 22100 Como, Italy}

\begin{document}

\begin{abstract}{
We provide a joint numerical-analytical study of the physics of a flowing atomic Bose-Einstein condensate in the combined presence of an external trap and a step potential which accelerates the atoms out of the condensate creating a pair of neighbouring black- and white- hole horizons. In particular, we focus on the rapidly growing density modulation pattern that appears in the supersonic region, an experimentally observed feature that was related to black-hole lasing phenomena. A direct assessment of the role of interactions in this process suggests an interpretation of the experimental data in terms of linear interference of atomic waves rather than collective effects. Our conclusions are further supported by an analytical solution of the Schro\"odinger equation in terms of Airy wavefunctions.}
\end{abstract}

\maketitle
\section{Introduction}

The observation and the characterization of the analogue Hawking emission \cite{hawking1,hawking2} from sonic black holes \cite{unruh} has been an intensive field of research in recent years \cite{liberati}. In particular, ground-breaking experimental results were reported in effectively one-dimensional (1D) flowing atomic Bose-Einstein condensates by J. Steinhauer at Technion~\cite{steinhauer1, steinhauer2,steinhauer3,steinhauer4}. These pioneering experiments have stimulated a number of refined theoretical and numerical studies to fully understand the physics at play there~\cite{numerical,ted1,ted2,steinhauer_pra,ulf,steinhauer_annalen,epl,prd,nonlinear, phon,finke,low,ll,iso,uni}. 

The goal of this work is to make another step in this direction by further investigating the first pioneering experiment in~\cite{steinhauer1}, where the observation of self-amplifying Hawking radiation in an analogue black-hole laser was first claimed. The general idea of black-hole lasing was first introduced in~\cite{ted} and then studied in more detail by several authors for different condensed matter and optical platforms in~\cite{renaud1,renaud2,renaud3,iac1,iac2,Faccio_2012,gaona2017theory}: in a stationary flow configuration displaying a pair of black- and white-hole horizons, the black-hole laser instability gives rise to complex-frequency eigenmodes, spatially concentrated in the supersonic region located between the horizons. Differently from standard lasing in optical cavities, amplification in a black-hole laser is provided by stimulated Hawking processes during the reflection at the horizons. In the experimental settings of~\cite{steinhauer1}, the identification of the microscopic processes at the origin of the observations is complicated by the time-dependence of the flow.

This has triggered intense discussions within the community~\cite{numerical,ted1,ted2,steinhauer_pra} on the interpretation of the experimental observations. In particular, several authors remarked that the temporally growing modulation in the experimental density profiles could be reproduced by mean-field Gross-Pitaevskii calculations, so they must have a classical hydrodynamical origin~\cite{numerical,ted1,ted2}.
While all these works agree in claiming that the spatial density modulation is triggered by Bogoliubov-Cerenkov emission in the supersonic flow region, it was pointed out in~\cite{ted1,ted2} that the temporal growth of the modulation pattern is not necessarily a signature of a classical hydrodynamic instability (as we had instead claimed in \cite{numerical}), but could also be explained more simply in terms of the temporally growing average density in the supersonic region. 

The goal of this article is to contribute to this debate by investigating the physics at play in the supersonic region. In particular, we will focus on the spatial modulation of the {\it average density}; all the specific features of the correlation function of {\it density fluctuations}, discussed in~\cite{steinhauer_pra,steinhauer4} and attributable to quantum fluctuations and Hawking processes, will not be addressed. In the supersonic region the atoms, initially accelerated by the waterfall potential at the (outer) black-hole (BH) horizon, get slowed down while climbing up the shallow trap potential, eventually forming a white-hole (WH) horizon. In particular, we focus our attention on the role played by inter-particle interactions in this region, where the atomic density is low and one can reasonably expect that the interactions between atoms play a minor role, at least in the early to intermediate time dynamics. Since the exponential growth of the density modulation was used as a possible signature of the onset of black-hole lasing and the mode conversion processes underlying black-hole lasing crucially depend on interactions, it is interesting to investigate the consequences of (artificially) switching off the interaction constant in the region beyond the sharp edge, while leaving the rest of the atomic cloud unaffected. 

\section{The physical system and the theoretical model}
\label{sec:model}

\begin{figure}[ht]
  \begin{center}
    \includegraphics[width=.6\linewidth]{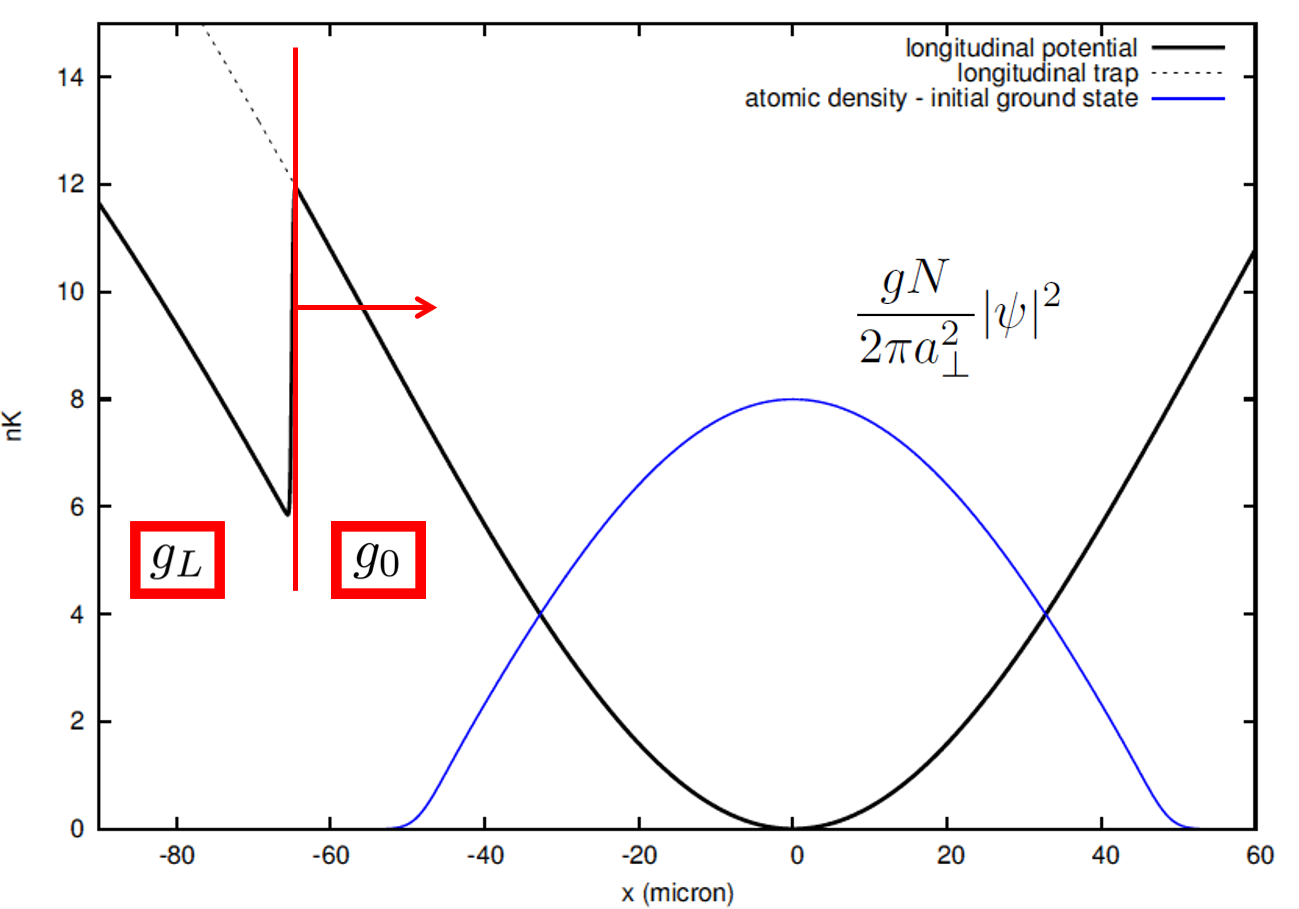}
      \caption{Sketch of the system under consideration: the ``waterfall'' step potential forming the black hole moves to the right with a constant speed $v$ across the initially trapped BEC. To assess the role of interactions, in our toy model the interaction constant to the right of the step potential is left at its physical value $g_0$, while it is suppressed to $g_L$ to the left of the step.}
      \label{trick2} 
  \end{center}
\end{figure}

\begin{figure*}[ht]
  \begin{center}
     \includegraphics[width=\textwidth]{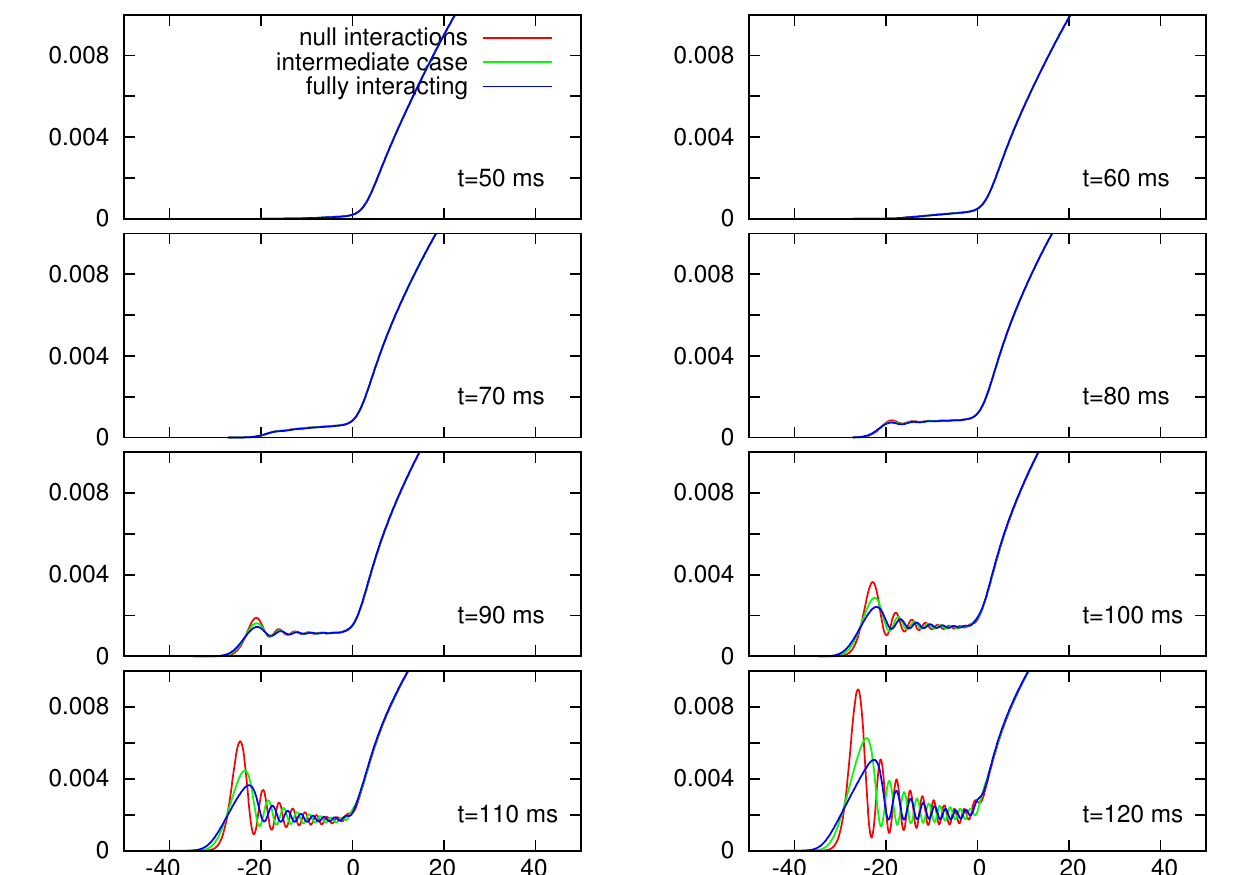} 
      \caption{The different panels show snapshots of the condensate density profile at subsequent times. The results are given in the waterfall potential step reference frame so that the black hole horizon is fixed at the origin while the BEC moves to the left. The $x$-axis unit is in $\mu$m while the $y$-axis one is in $\mu$m$^{-1}$. Within each panel, the three curves refer to different values of the interaction constant $g_L$ in the region to the left of the potential step: non-interacting $g_L=0$ case (red), intermediate $g_L=g_0/2$ (green), physical case $g_L=g_0$ (blue).}
      \label{profiles_snapshots} 
  \end{center}
\end{figure*}

In the experiment \cite{steinhauer1} the condensate is made of $^{87}$Rb atoms trapped in a confining laser beam that constrains the BEC cloud to have a elongated cigar shape and a nearly 1D dynamics. As it is shown in Fig. (\ref{trick2}), in addition to the confining trap a step-like optical potential is applied along the axial direction (i.e., the direction of motion) in order to accelerate the atoms and create the horizons. This ``waterfall'' potential is swept through the atomic cloud at a constant speed from the left to the right: in this way, as it is shown in the snapshots in Fig. (\ref{profiles_snapshots}), to the right of the potential step the condensate remains essentially unperturbed and at rest in the laboratory frame while to the left of the potential step the atoms acquire a supersonic speed in the leftward direction, forming the analogue of a black-hole horizon in the vicinity of the waterfall potential. Further away to the left of the BH, the leftward flow velocity decreases again as atoms climb the trap potential and approach the relevant turning point, where a density peak forms due to the accumulation of atoms. The point where the velocity of the atoms drops below the speed of sound is the white-hole horizon, which occurs before the atoms reach the turning point where their velocity vanishes. In this way, the experiment displays a pair of analogue BH and WH horizons enclosing a finite region of supersonic flow.

In order to model the experimental configuration, we describe the dynamics of the atomic cloud by means of an effective 1D Gross-Pitaevskii equation (GPE) \cite{dalfovo}:
\begin{equation}\label{gpe}
i \hbar \dfrac{\partial \psi}{\partial t} = \left( -\dfrac{\hbar^2}{2m}\dfrac{\partial^2 }{\partial x^2} + U(x) + \dfrac{g N}{2\pi a_{\bot}^2} |\psi|^2 + \hbar \omega_{\bot} \right) \psi \, .
\end{equation}
where $\psi(x,t)$ is the longitudinal part of the wave function while the transverse profile has been assumed to be fixed and Gaussian. Here, $a_{\bot}=\sqrt{\hbar/(m \omega_{\bot})}$ indicates the transverse size of the condensate in terms of the frequency $\omega_{\bot}$ of the harmonic trapping along the transverse $y,z$ directions; the atom-atom interaction constant $g$ is initially assumed to be spatially uniform and equal to $g_0=4\pi\hbar^2a_s/m$, with $a_s$ the atom-atom scattering length. This choice differs from the non-polynomial Schr\"odinger equation (NPSE) \cite{alb} used in our previous work but this technical simplification does not change the conclusions of our analysis.

$U(x)$ is the external trap potential in the longitudinal direction. For this latter, we choose a potential of the form
\begin{equation}\label{U0ted}
U{_{trap}}(x)=U_0 \, {x^2}/{(x^2+x_0^2)} \, ,
\end{equation}
that is plotted as a dashed line in Fig. (\ref{trick2}). This potential is obtained from the consideration that the constricting laser beam has a Gaussian shape, as done in \cite{ted1,ted2}. Note that this choice differs from our previous work \cite{numerical} where we had chosen a polynomial form fitting the experimental potential. Anyway, the precise form of the trap potential does not affect the phenomena examined in this work. 

The parameters of eqs. (\ref{gpe}) and (\ref{U0ted}) can be extracted from the specific experimental set-up. For a 3D Gaussian laser beam~\cite{steinhauer1}, we have that the potential is
\begin{equation}\label{laserbeam}
U_{trap}(r,x)=U_0 \, \left(1- \left[ \frac{w_0}{w(x)}\right]^2 e^{-\frac{2r^2}{w(x)^2}} \right) \, ,
\end{equation}
where $x$ is the longitudinal direction, $r=\sqrt{y^2+z^2}$ is the radial distance from the center of the beam, and the radial width $w(x)$ varies with the longitudinal coordinate as
\begin{equation}
w(x)=w_0 \, \sqrt{1+\left( {x}/{x_0}\right)^2 }\ \, .
\end{equation}
Here, $w_0$ is the beam waist and the Rayleigh length $x_0$ depends on the laser wavelength $\lambda$ as $x_0={\pi w_0^2}/{\lambda}$. Along the axis $y=z=0$, this form recovers exactly Eq. (\ref{U0ted}). Furthermore, if we expand at the lowest order the trap potential (\ref{laserbeam}) around $r=0$ we have
\begin{equation}\label{omeg}
U{_{trap}}(r,x) \simeq -\frac{U_0 w_0^2}{w(x)^2}+\frac{2 U_0 w_0^2}{w(x)^4} \, r^2 = -\frac{U_0 w_0^2}{w(x)^2}+\frac{1}{2} m \omega^2_\perp(x) \, r^2 ,
\end{equation}
which gives, in the center of the trap, $U_0=(1/4)m\,\omega_{\bot}^2w_0^2$. Thus, given the experimental values $w_0=5 \, \mu$m, $\lambda=812$ nm and $\omega_{\bot}=\omega_{\bot}(0)=(2\pi)\,123$~Hz \cite{steinhauer1}, the transverse trap dimension is $a_\perp=\sqrt{{\hbar}/{m\omega_\perp}}=0.97\,\mu$m. 

The zero-point motion along the transverse $y,z$ directions gives the last term of Eq. (\ref{gpe}). In the following, we shall make the further simplification of neglecting the longitudinal $x$-dependence of $\omega_\bot$. Even though such an approximation may have a significant quantitative impact on the interaction constant and the total longitudinal potential experienced by the atoms, it does not affect the qualitative conclusions of our work, as can be easily demonstrated by numerical means.

Including also the step-like potential which accelerates the atoms to supersonic speeds and creates the horizons, the total potential (solid line in Fig. (\ref{trick2})) turns out to have the form:
\begin{equation}\label{axialpotential}
U(x,t)=U_0 \, \frac{x^2}{x^2+x_0^2}-\frac{U_s}{2}\left[1-\tanh\left(\dfrac{x-vt-x_s}{\sigma}\right)\right] ,
\end{equation}
where $U_s=6.4$ nK, $v=0.21$ mm/s, $\sigma=2 \, \mu$m are the experimental parameters of the waterfall potential \cite{steinhauer1} and the initial position $x_s\ll 0$ is chosen so that at the initial time $t=0$ the step is far from to the left of the cloud.

Throughout this work, we integrate Eq. (\ref{gpe}) using as the initial condition the ground state of the system in the trap potential, calculated via imaginary-time evolution with the experimental value of the chemical potential $\mu$ shifted\footnote{Note that the value $\mu=8$~nK quoted in~\cite{steinhauer1} does not include the transverse zero-point energy term in Eq. (\ref{gpe}).} by the zero-point energy $\hbar \omega_{\bot}$.

\section{Numerical results: the role of interactions}
\label{sec:num}

From both the simulations \cite{numerical,ted1,ted2,steinhauer_pra} and the experimental data \cite{steinhauer1}, it is evident that the density is very low in the region to the left of the waterfall potential (see, for example, Figs. (2) j-r of \cite{steinhauer1}). This is easily explained: initially, the waterfall potential is located far away from the condensate [first panel of Fig. (\ref{profiles_snapshots})]. As time passes, the waterfall potential approaches the center of the condensate, so an increasing amount of atoms starts to flow over it (second to fifth panel) and to gain velocity. Due to number conservation, such an increase in the velocity implies a corresponding decrease in the density and, thus, on the left of the potential drop, we have a reduced density of fast moving atoms. This simple observation led us to wonder if inter-particle interactions, which are a necessary ingredient for collective effects like the black-hole laser mechanism~\cite{renaud1,renaud2,renaud3,iac1,iac2}, play an important role in the dynamics or they could, in principle, be neglected. 

To investigate this issue, we can introduce a toy model where the strength of the interaction is tuned in a position-depending way as $g(x)$: it is left at its unperturbed value $g_0$ to the right of the step and is suppressed to the left of the step to $g_L$, as represented in Fig. (\ref{trick2}). In formula, this is included via an interaction constant of the form\footnote{The smoothness of the transition between $g_0$ and $g_L$ - regulated here by the longitudinal trap parameter $\sigma=2 \, \mu$m - does not affect the conclusions of this work.}:
\begin{equation}
g(x)=\frac{g_0+g_L}{2}+\frac{g_0-g_L}{2} \tanh\left(\dfrac{x-vt-x_s}{\sigma}\right) \, .
\label{toy}
\end{equation}
Thus, we numerically integrate this modified GPE starting from the ground state of the interacting condensate with the original interaction constant $g_0$ and, when the atoms encounter the potential drop, interactions are also switched off to $g_L=0$. This has the effect of transforming Eq. (\ref{gpe}) in the well-known Schr\"{o}dinger equation to the left of the waterfall, while to the right the full GPE keeps being integrated. Rather than integrating the non-interacting Schr\"{o}dinger equation on the whole axis, we chose this strategy to ensure an accurate description of the condensate in the trap region, where the density is largest. On the other hand, collective effects in the \textit{supersonic} region are completely disregarded. As a consequence, all the features observed to the left of the sonic horizon just reflect the trivial interference effects which characterize a linear wave packet dynamics.

The result of the numerical integration of the GPE (\ref{gpe}) for this toy model is plotted in Fig. (\ref{profiles_snapshots}) where the evolution of the density profile is shown in three different cases. For the sake of clarity, we plot the results in the reference frame of the potential step, so that the BH horizon is at rest at the origin and the condensate moves to the left. In all three cases, to the right of the potential step we have the fully-interacting cloud with $g_0$, while different values of the interaction parameter $g_L$ to the left of the potential step are used. In the specific, the numerical integration of the full GPE with $g_L=g_0$ is shown as a blue curve, while the result of the toy model with $g_L=0$ are displayed as a red curve. In order to better understand the transition between these two cases, we also integrate the toy model with an intermediate value $g_L=g_0/2$ (green line).

\begin{figure}[ht]
  \begin{center}
      \includegraphics[width=.6\linewidth]{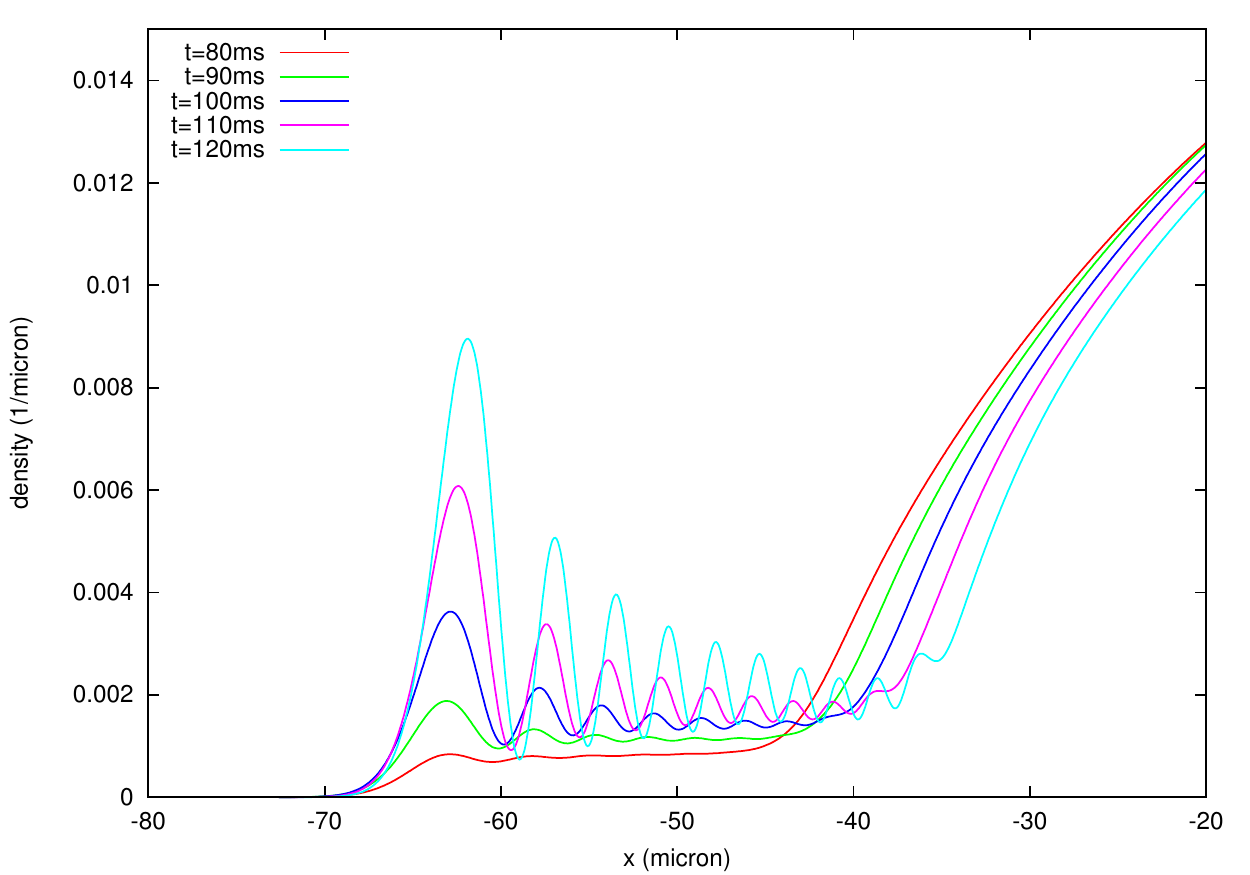}
      \caption{Time evolution of the density profile as seen from the trap reference frame. The different curves are taken at five times $t=80, 90, 100, 110, 120$~ms for the same parameters as the red curves of Fig. (\ref{profiles_snapshots}), in particular a vanishing interaction constant $g_L=0$ on the left of the potential step.}
      \label{schro} 
  \end{center}
\end{figure}

\section{Interpretation of the numerical results}
\label{sec:results}

Fig. (\ref{profiles_snapshots}) shows the dynamics of the atomic cloud at eight subsequent times. In the reference frame of the potential step, the BEC moves from the right to the left with a constant speed: the particles start in the interacting ground state in trap potential and, at a certain time, they encounter the potential drop, which accelerates them. As time passes, an oscillatory pattern is seen to appear in the density profile in the region to the left of the step and the contrast of these fringes appears to grow in time.

This is true for all the three cases plotted in the figures which represent, to the left of the step, the case of vanishing interactions $g_L=0$, the one of a fully interacting cloud $g_L=g_0$ and the intermediate case with $g_L=g_0/2$. The figure shows that there is a smooth transition from one case to the other. In particular, the total length of the oscillatory pattern in the density profile as well as the number of peaks remain approximately the same and they agree with previous numerical simulations and with the experimental results~\cite{steinhauer1,numerical,ted1,ted2,steinhauer_pra}. As the main differences, the presence of repulsive interactions results in a slight shift of the effective turning point in the total potential determined by the trap and the interactions and in a broadening of the interference peaks in the density.

The fact that the same pattern is retrieved in all three cases regardless of the strength of interactions to the left of the potential step (i.e., even when they are absent) is the main result of this paper, which pushes us to two conclusions. First of all, it confirms that the fringe pattern observed in the density profile in \cite{steinhauer1} has a purely classical interpretation: that is, zero-point quantum fluctuations of the phonon field do not a play a significant role in this feature, confirming the conclusions of ~\cite{numerical,ted1,ted2}. Secondly, even more importantly, given that no dynamical instability can take place in Schr\"odinger-like linear differential equations, our calculations suggest that the temporal growth of the fringe pattern can hardly be attributed to collective instability effects such as black-hole lasing and, confirming the claim in~\cite{steinhauer_pra}, is not necessarily related to the presence of a WH horizon. The appearance of the fringes can in fact be explained in terms of a linear matter-wave interference phenomenon between the atoms that are reflected by the trap potential and the ones that keep being accelerated by the potential step, with minor modifications due to interactions. In contrast to collective instability mechanisms and to related statements in our previous work~\cite{numerical}, but in agreement with the claim of~\cite{ted1,ted2}, the rapid growth of the fringe pattern can be thus traced back to the steep density profile of the original condensate, that induces a rapid increase of the particle flux that is injected into the ``supersonic'' region by the condensate hitting the waterfall potential. In this context, we wish to stress that the Bogoliubov-Cherenkov emission mechanism~\cite{bogcer}  put forward in~\cite{numerical,ted1,ted2} continuously transforms into a standard reflection process when interactions are brought to zero, so our results do not contradict the works in~\cite{ted1,ted2}, but rather support them. 

Of course, all our discussion is restricted to early and intermediate times after the formation of the BH horizon, namely well before the right-moving reflected atoms have had time to return the BH. At this point, as it has been experimentally investigated in~\cite{steinhauer4}, collective effects, possibly related to stimulated Hawking processes, could start to play a role, providing an effective amplification mechanism. But this is beyond the scope of our analysis, which is focused on the temporal growth at early times when such processes are not yet active.

\section{Analytical understanding of the dynamics}
\label{sec:anal}

Having numerically shown that interactions do not play a major role in the early dynamics in the region to the left of the potential step, we can exploit the mathematical simplicity of the Schr\"odinger equation to search for an analytical explanation in terms of simple wavefunctions. 

In Fig. (\ref{schro}) we re-plot the red curves of Fig. (\ref{profiles_snapshots}) for a vanishing interaction to the left of the step, now superimposed in the reference frame of the trap where the potential step moves to the right. Here, we can see that the fringes grow in time and they slowly move to the right  (slower than the potential step), as observed in both the experiment\cite{steinhauer1} and in the numerics \cite{numerical,ted1,ted2,steinhauer_pra}. Specifically, our task will be to analytically recover both these effects.
\begin{figure} [ht!]
\includegraphics[width=.6\linewidth]{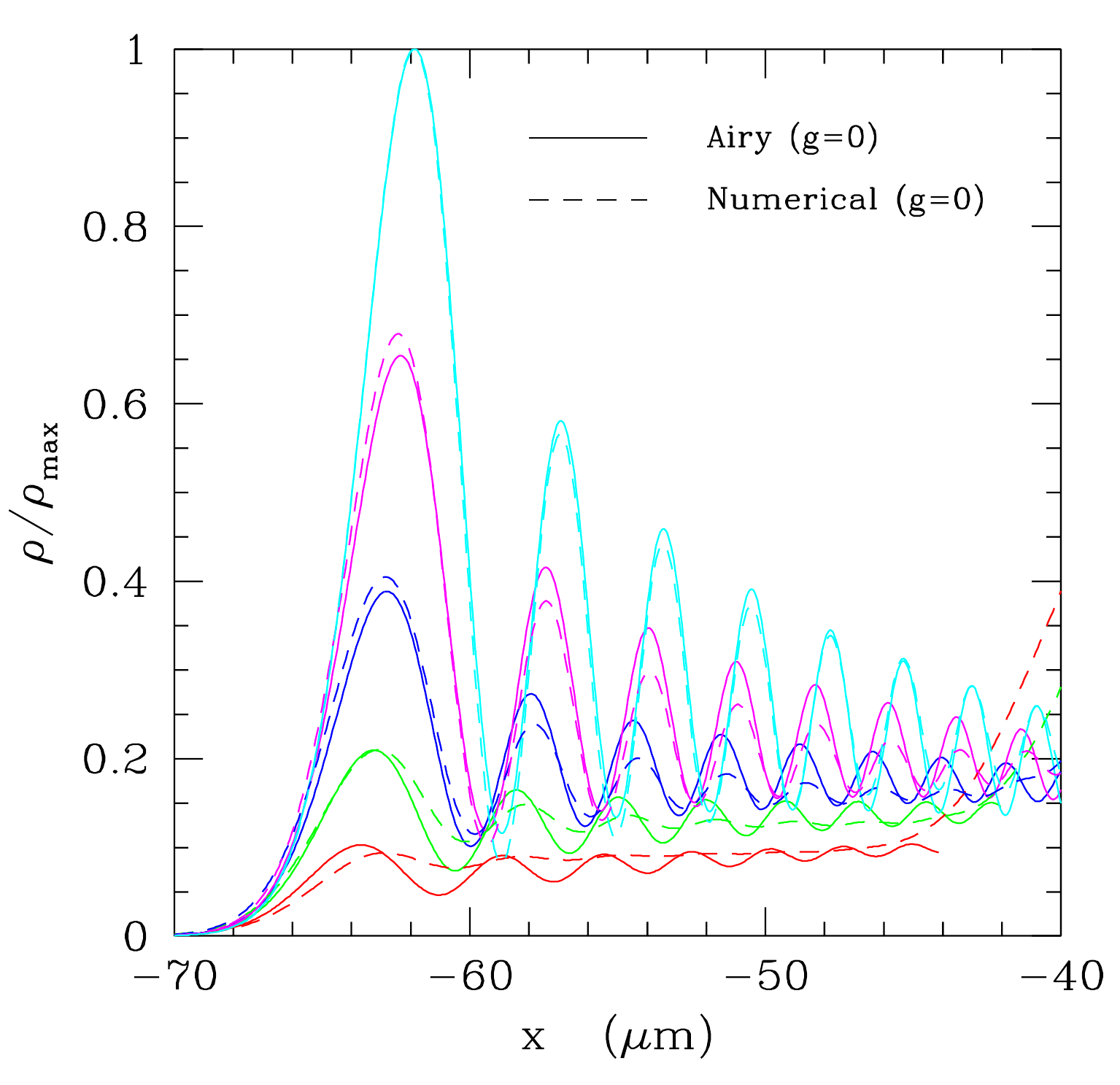}
\caption{The solid lines represent the density profile of the analytical wavefunction Eq. (\ref{sol}) while the dashed lines are the numerical solutions for $g_L=0$. The different colors correspond to five representative times $t=80, 90, 100, 110, 120$~ms.}
\label{fig1}
\end{figure}

To do so, we start by linearising the trap potential $U(x)$ around the ``turning point'' $x=\bar x$ defined by 
$U(\bar x) = \mu=8$ nK: $U(x) \sim \mu -\alpha(x-\bar x)$ with $\bar x =-49.27 \, \mu$m  and $\alpha =0.258$ nK/$\mu$m. Then, we perform a change of reference frame setting $X=x-vt-x_s$ so that the step is now located at $X=0$ at all times. The Schr\"odinger equation in the region $X < 0$ then becomes \footnote{The correct approximation would be to consider $X \ll \sigma$ but here, for the sake of simplicity, we consider a sharp step, i.e., $\sigma \to 0$.}
\begin{equation}\label{schrodinger}
i \hbar \left [ \frac{\partial\psi(X,t)}{\partial t} -v\,\frac{\partial\psi(X,t)}{\partial X}\right ] 
= -\frac{\hbar^2}{2m}\frac{\partial^2\psi(X,t)}{\partial X^2} + \left [\mu -U_s-\alpha(X+vt+x_s-\bar x)\right ]\, \psi(X,t) \, .
\end{equation}
Finally, we look for a factorized solution in the range $X\in [-\infty,0]$: $\psi(X,t)=f(t)\,g(X+wt)$ where $w$ is an unspecified, generally complex, number. Defining $\xi=X+wt$ and substituting into (\ref{schrodinger}) we obtain 
\begin{equation}
g(\xi)\left \{i \hbar \frac{df(t)}{d t} -\left [ \mu-U_s+\alpha(\bar x -x_s+(w-v)t)\right ]\,f(t) \right \}= 
f(t)\,\left \{i \hbar\,(v-w)\,\frac{dg(\xi)}{d\xi} -\frac{\hbar^2}{2m}\frac{d^2g(\xi)}{d \xi^2} - \alpha\,\xi\, g(\xi) \right \} \, .
\end{equation}
Dividing by $f(t)g(\xi)$ and equating each side of the equation to an arbitrary common (complex) constant $\epsilon$, we obtain two independent equations for $f(t)$ and $g(\xi)$. Setting 
\begin{equation}
g(\xi) = e^{-i \frac{m}{\hbar}\,(w-v)\,\xi} A(\xi) 
\end{equation}
the function $A(\xi)$ obeys the Airy equation and is therefore given by 
\begin{equation}
A(\xi) = Ai\left ( -\left [\frac{2m\alpha}{\hbar^2} \right ]^{1/3} \left [ \xi +\frac{m (w-v)^2+2\,\epsilon}{2\,\alpha}\right ] \right ) 
\end{equation}
while $f(t)$ is an exponential function,
\begin{equation}
f(t) = e^{-\frac{i}{\hbar}\left [ (\epsilon  +\mu-U_s+\alpha\,(\bar x-x_s))\,t+\frac{\alpha}{2}\,(w-v)\,t^2\right ] \, .
}
\end{equation}
Collecting all terms and expressing the solution in terms of the original variables we finally get 
\begin{equation}\label{sol}
\psi(x,t) \propto \text{exp} \left( -\frac{i}{\hbar}\left [ (\epsilon +m\,\bar w^2)\,t +\frac{\alpha}{2}\,\bar w\,t^2+m\,\bar w\,x\right ] \right) \textrm{Ai} \left ( -\left [\frac{2m\alpha}{\hbar^2} \right ]^{1/3} \left [ x-x_s+\bar w\,t+\frac{m \bar w^2+2\,\epsilon}{2\,\alpha}\right ] \right ) 
\end{equation}
where $\bar w = w-v$ and a constant pre-factor, including a time dependent phase factor, has been dropped. Note that the first exponential is not necessarily a phase factor because the two integration constants $\epsilon$ and $\bar w$ 
may include an imaginary part.
Of course, Eq. (\ref{sol}) is a particular class of solutions of the Schr\"odinger equation and the general solution is obtained by an arbitrary superposition of these forms. Nevertheless, we can hope that, by suitably choosing the two constants ($\epsilon, \bar w$), a good representation of the numerical integration can be reached with a single wavefunction. 

\begin{figure} [ht!]
\includegraphics[width=.6\linewidth]{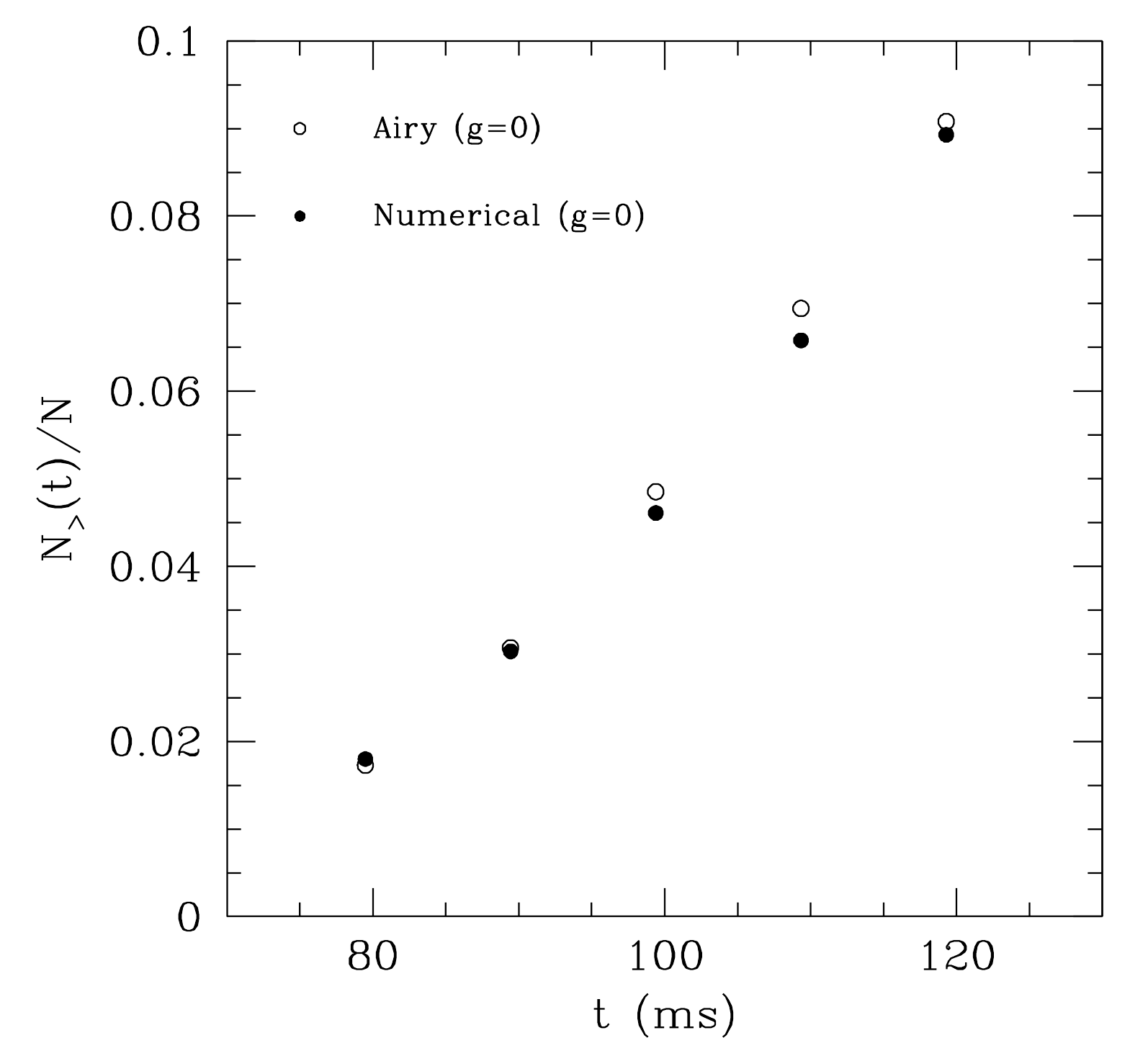}
\caption{The full dots show the time-dependence of the number of particles on the left of the step extracted from the numerical simulation for $g_L=0$, while open black dots are obtained by integrating the modulus square of the analytical expression (\ref{sol}) for $x < x_s +vt$. The time-independent normalization constant (unspecified in Eq. (\ref{sol})) has been fixed by matching the height of the first peak in the density profile with the exact GPE result at $t=120$ ms (as done in Fig. (4)).
}
\label{fig2}
\end{figure}
The results are shown in Fig. (\ref{fig1}): the simple solution (\ref{sol}) turns out to be already a good approximation of the numerical result. Indeed, the Airy wave resembles the numerical data and the peaks are seen to grow and move in time, as in the experiment \cite{steinhauer1} and the simulations \cite{numerical,ted1,ted2,steinhauer_pra}. Note that the analytical solution (\ref{sol}) does not need to be  norm conserving: due to the motion of the step potential, the number of particles entering the region of free time evolution is in fact steadily increasing. Because of this peculiar boundary condition, we are entitled to take the constant $w$ and $\epsilon$ as possibly complex. In particular, the values corresponding to Fig.(\ref{fig1}) are 
\begin{align} 
\bar w = \left (-0.052 -0.016\,i \right)\, \text{mm/s} \\ \nonumber 
\frac{m\,\bar w^2+2\,\epsilon}{2\,\alpha}= \left( 5.36 +2.54\,i \right)\, \mu\text{m}.
\end{align}
As one can see in the figure, this choice allows to accurately recover the temporal evolution of the fringes observed in the experiments and in the numerics, in particular their growth and their slow motion.

More specifically, if we had $w=0$ (i.e. $\bar{w}=-v$) the peaks would have stayed stationary in the reference frame of the step. In \cite{ted1,ted2} the motion of the fringes in the BH frame was attributed to the movement of the WH which rigidly drags the zero-frequency Bogoliubov-Cerenkov emission. Here, the difference in velocities in the Airy solution already accounts for this ``Doppler shift''. It can be shown by means of classical calculations that this discrepancy traces back to the fact that, since the longitudinal potential is time-dependent, energy is not conserved. This consequent temporal variation of the kinetic energy of the accelerated atoms makes the turning point to slightly move during the evolution. Since the position of the fringe pattern is fixed by the one of the turning point, the motion of this latter causes a corresponding drift in the density peaks.

Also, most remarkably, a non-vanishing imaginary part in the parameters allows the peaks to grow in time according to an exponential law even in the absence of interactions, at least at small times. This behaviour was investigated in \cite{steinhauer1} and debated in \cite{numerical,ted1,ted2,steinhauer_pra}. Since no dynamical instability can take place in a linear equation such as the Schr\"odinger one, the fact that the main features of the density pattern are qualitatively recovered by our Airy ansatz with $g_L=0$ rules out a major role of collective instabilities such as black-hole lasing in determining the growth of the density modulation pattern at early and intermediate times. Of course, the accuracy of this analytical ansatz breaks down at later times when the dynamics is complicated by the scattering of the reflected wave onto the original condensate.

Further insight on this physics is given in Fig. (\ref{fig2}), which shows the fast temporal growth of the number of particles to the left of the step potential for the non-interacting case as compared to the Airy function model. Once again, the agreement of the two sets of points indicates that the dynamics to the left of the potential step can be explained in terms of a linear Schr\"odinger equation with an Airy-function-shaped ansatz. The minor quantitative differences between the analytical solution (\ref{sol}) and the numerical results - also evident in Fig. (\ref{fig1}) - are due to the fact that we are considering a single Airy wavefunction, while the linearity of the Schr\"odinger equation allows the wave function to be a linear superposition of several of them. 

\section{Conclusions}
\label{sec:conclu}

In this paper we have developed a simple model to understand the origin of the temporally growing density modulation pattern that was observed in a recent experiment claiming black-hole lasing~\cite{steinhauer1}. By performing Gross-Pitaevskii numerical simulations for the condensate dynamics, we have shown that interactions in the spatial region past the black-hole horizon are not crucial to the onset of the density modulation and that the qualitative features are unaffected when interactions in this spatial region are switched off. This suggests that the presence of a white-hole horizon and the Hawking amplification effects at the horizons are not playing a central role in the dynamics of the average density profile, the observed modulation pattern being rather due to a linear interference between the atoms that get reflected by the trap and the ones that keep being accelerated out of the original BEC by the potential step. In particular, in contrast to our previous claims in~\cite{numerical} and in agreement with~\cite{ted1,ted2}, the temporal growth of the pattern can be simply explained in terms of the steep density profile of the original condensate hitting the potential step.

As a further confirmation of our conclusions, we have found an analytical solution of the Schr\"odinger equation based on Airy functions that qualitatively recovers the main features of the numerical solution. Since Schr\"odinger-like linear equations can not display dynamical instabilities, this rules out collective effects such as black-hole lasing as a possible mechanism to explain the rapid onset of the density modulation at early times. The situation is of course very different for the later time dynamics of fluctuations on top of the Gross-Pitaevskii solution that was addressed in the latest experiments~\cite{steinhauer4}.

\section*{Acknowledgements}

We dedicate this work to the memory of our friend and master Renaud Parentani with whom this work was developed along many years of fruitful collaboration. If it were not for his premature departure, his name would have been among the authors. All credit for our achievements must be shared with him, all discredit for mistakes is fully the three authors'.

%%%%%%%%%%%%%%%%%%%%%%%%
\bibliographystyle{ieeetr}
\bibliography{Bibliography}

\end{document}